# The Age-Competency Model to the Study of the Age-Wage Profiles for Workers

*Our main need is neither food, nor money. It is education.*

Serguei I. Maximov

*Republic of Belarus National Institute for Higher Education*

In this article, I present a new approach and a novel model to the study of the life cycle of wages. The key idea is that wage can be thought of as remuneration paid for the competency. It is assumed with the approach that there are three main mechanisms acting at micro level and resulting in the change of workers' competencies during their lives. These are an endogenous growth of workers' initial competencies; a rate of investment in schooling in the life cycle of wages; and an effect of relative losses in workers' competencies. The developed macro model is to shed light on the processes resulting in the age-wage profiles seen in mass. The model obeys a nonlinear integro-differential equation. The found analytic solution of the equation has the form of Fisk PDF of a special type. The solution and its features are discussed. The regression technique is used to check the model upon reliability. The model provides better fitting to the data (Elo & Salonen, 2004) than minceranian earnings function (Mincer, 1974) does.

## Introduction

The study of the effects of investment in schooling and on-the-job training on the level, pattern and interpersonal distribution of life-cycle earnings have been pioneered by Becker (1962, 1964, 1965, 1966) and Mincer (1958, 1962, 1974). This leads to one of the most successful empirical equation called the "Mincer (1974) earnings function"

$$\ln(w) = \beta_0 + rS + \beta_1 E + \beta_2 S^2 + \varepsilon,$$

where $w$ is earnings (or the hourly wage when available), $S$ is years of schooling, $E$ is years of labor market experience, and $\varepsilon$ is an error term. Years of schooling are a fairly direct measure of educational human capital, while years of labor market experience are viewed as a proxy for on-the-job training.

Almost all empirical studies find that schooling has a positive and significant effect on earnings ($r > 0$), these earnings are a concave function of labor market experience ($\beta_1 > 0$ and $\beta_2 < 0$). The Mincer equation thus captures the important empirical regularities:

    1) increase in earnings with schooling,
    2) concavity of log earnings in experience,
    3) parallelism in log earnings experiences profiles for different educational groups (ratio of earnings for persons with education levels differing by a fixed number of years is roughly constant across schooling levels)
    4) U-shaped interpersonal variance in earnings.

The theoretical foundations of Mincer specification can arise from two theoretical frameworks:
    i. compensating wage differentials (Mincer, 1958)
    ii. accounting identity framework (Mincer, 1974).

Technically, the Mincer's approach is based on the idea of allocation of non-leisure time between working and schooling. Namely, to get the Mincer model it should be assumed that the workers used to share their non-leisure time between schooling and working. Such an assumption looks very natural, but there is a problem how to separate and to measure these pieces of time, and especially in on-the-job training.

The basic Mincer human capital earnings function does not appear to fit the data nearly as well in the 1980s and 1990s as it did in the 1960s and 1970s:

> 1) Log wages are an increasingly convex function of years of schooling: that is, the log-linearity in schooling seems to be held no longer.
> 2) The quadratic in experience tends to understate earnings growth at the beginning of the lifecycle and overstate the decline in earnings towards the end of the lifecycle. Murphy and Welch (1990) find that a quartic fits the data better.
> 3) Experience-wage profiles are no longer parallel for different educational groups: that is, the multiplicative separability between experience and schooling holds no longer. For example, the college-high school wage gap is now much larger for less experienced than for more experienced workers.

It is very likely the investments in human capital, the workers in mass made in recent decades for their earnings to grow up, have had the determining effect on their wages.

In this article, I present a model to explain the effect. To build the model I have developed a novel approach, which is not a "clone" of Mincer's; the approach has nothing of the idea of allocation of a non-leisure time between working and schooling.

**The framework of approach**

Worker's competency is a key concept I apply to in my approach. The common meaning of noun "competency" is the quality of human being adequately and well qualified physically and intellectually for doing a certain job. Schooling is the only way for people to become well qualified intellectually. It can be considered the law, which is acting in the world of education – the more schooling is taken, the more level of competency is gotten.

I assume the workers' competencies are the only goods they sell on the labor market. This means that the competencies have some market values, and wages of workers are "made" of these. It is seen and can be considered the law, which is acting in the world of economy, that the higher level of competency brings as a rule the higher wages.

Worker's competency tends to vary. Extra schooling and on-the-job training boost the worker's competency. In my approach, I pre-assume an extra schooling is paid for; to get more level of competency the worker invests in himself some part of the personal income. Such pattern of economic behavior of workers looks quite natural; one can expect they in mass do so to have higher wages. It is quite natural to assume also that there is an endogenous growth of the workers' competencies as a result of inner development of knowledge and professional skills acquired by the workers. For brevity, I call this process "the endogenous growth of initial competency". The workers' competencies tend also to get out-of-date i.e. to go down, as the workers get older. Producing of new knowledge goes on constantly in the modern world; this knowledge is used to provide schooling; those who just have finished schools enter labor market with this new knowledge. The old workers in mass know as a rule a little of that new and they cannot become younger to get back to schooling from the very beginning. Accumulating knowledge makes human less flexible in acquiring new knowledge. Such accumulating goes on during the life cycle of human so people become the "hostages" of that knowledge that they have acquired. I propose a novel model for that, namely I assume that the specific rate of change of current competency tends to diminish with time-averaged accumulating competency. Formally, in my approach, this model is a substitute for that part of modeling in the field of economic of education that is known as the problem of overlapping generations. For short, the process of relative loosing of the worker's competency I term as "competency damping".

## The model

The basic model relied on the approach comes with the integro-differential equation

$$\frac{1}{C(t)} \cdot \frac{dC(t)}{dt} = -\alpha \cdot \frac{1}{t} \cdot \int_0^t C(\tau)d\tau + k_{\overline{C}} \cdot \frac{1}{t} + \beta \cdot \frac{1}{W(t)} \frac{dW(t)}{dt}, \qquad (1)$$

where $\alpha, k_{\overline{C}}, \beta > 0$ are some control parameters, and $C(t)$ and $W(t)$ denote age-competency and age-wage profile correspondingly; both $C(t)$ and $W(t)$ are taken to be positive and non-zero functions in the (workers') lifetime interval $[0, T]$ to be measured in years. Here, $T = Age_{getoff} - Age_{enter}$, and its regular value (in years) is about 40 as the workers in mass enter labor markets at $Age_{enter} \cong 20$ and then get off labor markets (due to attaining pension age) at $Age_{getoff} = 60$.

The left-hand side member of Eq. (1) is the specific rate of change of worker's competency; those in the right-hand side describe, correspondingly, competency damping, endogenous growth of initial competency, and competency growth due to investing some part of income in extra schooling.

I assume that the wages are linear with the competencies so

$$W(t) = \gamma \cdot C(t), \qquad (2)$$

where $\gamma$ is a constant.

By substituting (2) in (1) and then simplifying Eq. (1), I get

$$\frac{d \ln C(t)}{dt} = -A \cdot \frac{1}{t} \cdot \int_0^t C(\tau)d\tau + \frac{k_{\overline{C}}}{1-\beta} \frac{1}{t} \qquad (3)$$

where $A = \alpha/(1-\beta)$. (I have to point here that the solution of Eq. (3) has an economic sense if the component parameter, $A$ is a positive number.)

By making substitution $\ln C(t) = Z(t)$ in Eq. (3) one can re-write it in the form of

$$\frac{dZ(t)}{dt} = -A \cdot \frac{1}{t} \cdot \int_0^t e^{Z(\tau)} d\tau + B \frac{1}{t}, \qquad (4)$$

where $B = k_{\overline{C}}/(1-\beta)$.

The integro-differential equation (4) can be re-written then as a second-order ODE

$$\frac{d^2 Z(t)}{dt^2} + \frac{1}{t} \cdot \frac{dZ(t)}{dt} + \frac{A}{t} \cdot e^{Z(t)} = 0. \qquad (5)$$

## The analytic solution

The second-order ODE, Eq. (5) is a non-linear one but with linear symmetries that allows getting the exact analytic solutions (Polyanin & Zaitsev, 2003). The solution can be written in the form of

$$Z(t) = \ln\left( \frac{2P^2}{A \cdot Q} \frac{t^{P-1}}{\left[1+\left(\frac{t}{Q^{\frac{1}{P}}}\right)^P\right]^2} \right), \tag{6}$$

where

$$P = \sqrt{(1+Z'(1))^2 + 2A \cdot e^{Z(1)}}, \tag{6a}$$

$$Q = \frac{P^2 + (1+Z'(1))P - A \cdot e^{Z(1)}}{A \cdot e^{Z(1)}}, \tag{6b}$$

$$Z'(1) = \left.\frac{dZ(t)}{dt}\right|_{t=1} = \frac{1}{C(1)}\left.\frac{dC}{dt}\right|_{t=1}. \tag{6c}$$

Here, $Z'(1) > 0$ and $A > 0$ are assumed.

Thus, I get the formula for the age-competency profile,

$$C(t) = \frac{2P^2}{A \cdot Q} \frac{t^{P-1}}{\left[1+\left(\frac{t}{Q^{\frac{1}{P}}}\right)^P\right]^2}, \tag{7}$$

and also the formula for the age-wage profile (see formula (2)),

$$W(t) = \gamma \cdot \left( \frac{2P^2}{A \cdot Q} \frac{t^{P-1}}{\left[1+\left(\frac{t}{Q^{\frac{1}{P}}}\right)^P\right]^2} \right). \tag{8}$$

Both profiles are bell-shaped and have some fattened right tails; at the beginning of the life cycle of wages (or competencies), the curves go up slowly, reaching the flattened peaks, and then more slowly go down. There are only three controllable component parameters $A$, $P$, and $Q$ in the found solution. By varying these, one can modify the curves' shapes. These parameters depend on the model's parameters $\alpha$, $\beta$, and also on $Z(1)$ and $Z'(1)$ – the initial conditions set up with Eq. (3). It looks very natural to set up the initial conditions at $t = 1$, i.e. at the end of the first year at work, as at this moment both earnings and competencies of workers can be fixed (measured) for the first time.

These typical age-competency and age-wage profiles are shown in Fig. 1, 2 just below.

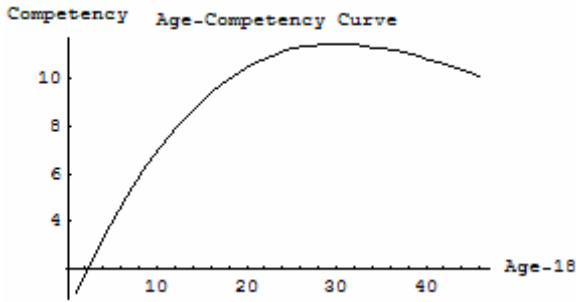
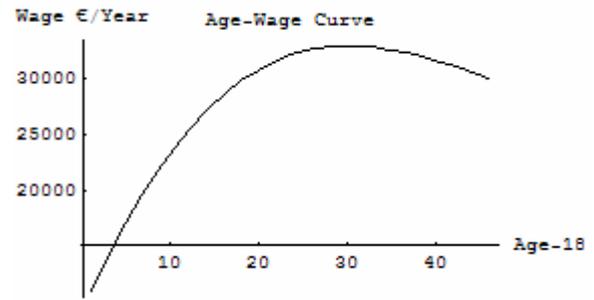

**Fig. 1.** The age-competency profile ($A=0.00364$, $P=1.872$, $Q=1925.3$)

**Fig. 2.** The age-wage profile ($A=0.00364$, $P=1.872$, $Q=1925.3$, $\gamma=2096.3$)

It follows immediately from Eq. (7) that both workers' competencies and wages peak at the age of

$$t_{max} = \left[ Q \frac{P-1}{P+1} \right]^{\frac{1}{P}}. \tag{9}$$

At this age, worker's competency is

$$C_{max} = C(t_{max}) = \frac{P^2-1}{2A} \cdot \frac{1}{t_{max}}. \tag{10}$$

Therewith, the magnitude of

$$C_{max} \cdot t_{max} = \frac{P^2-1}{2A} \tag{11}$$

does not depend on the worker's age; this magnitude can be considered an invariant of the model.

The wages also peak at the age of $t_{max}$, and the wages' maximum magnitude is

$$W_{max} = W(t_{max}) = \gamma \cdot \left( \frac{P^2-1}{2A} \cdot \frac{1}{t_{max}} \right). \tag{12}$$

I have to point out here that the age-competency profile (7) being formally considered as a distribution on $t \in [0,\infty\}$ represents a special case of the Fisk distribution. (The Fisk distribution is known also as loglogistic distribution, which is often used in income and lifetime analysis. In our case

$$\int_0^\infty C(t)dt = \frac{2P}{A}, \tag{13}$$

and so one can write

$$C(t) = \frac{2P}{A} \cdot \left\{ \frac{P}{Q^{\frac{1}{P}}} \cdot \frac{\left(\frac{t}{Q^{\frac{1}{P}}}\right)^{P-1}}{\left[1 + \left(\frac{t}{Q^{\frac{1}{P}}}\right)^{P}\right]^{2}} \right\}. \tag{14}$$

Here, in (14) the term in the figured brackets has the standard form of Fisk PDF (McLaughlin, 2001) but of specific kind. (Note that $Q$ depends on $P$ in this case (see formulas (6a), (6b)).

Thus, the solution of nonlinear Eq. (5) obeys Fisk PDF of specific kind (14). This is an important result of my study. Up to now, researchers look upon Fisk PDF as function, which is a suitable one for income data fitting. Now, relying on the model, Fisk PDF can be very naturally used for that.

It can be easily found by putting solution (7) in Eq. (3) that

$$P - 1 = \frac{k_{\bar{C}}}{1-\beta} = B. \tag{15}$$

Here, I have just rewritten Eq. (3) with using of (15), (6a) and (6c):

$$\frac{1}{C(t)} \frac{dC(t)}{dt} = -A \cdot \frac{1}{t} \cdot \int_0^t C(\tau) d\tau + \left[ \sqrt{\left(1 + \frac{1}{C(1)} \frac{dC}{dt}\bigg|_{t=1}\right)^2 + 2A \cdot C(1)} - 1 \right] \cdot \frac{1}{t}. \tag{16}$$

I call Eq. (16) the age-competency master-equation. It "rules" the processes having been discussed in the framework of my approach.

In the light of the approach, the coefficient at the second term in the right-hand side of Eq. (16) appears to be very natural; endogenous growth's trajectory of initial competency "starts" at some "right" angle at $t = 1$. Actually, if condition $Z'(1) \gg 2A$ holds, the starting slope of the endogenous growth's curve at $t = 1$ is very close to the slope of overall solution's curve. The condition means simply that the overall solution's trajectory will keep going for a while closely with the trajectory, which is predetermined by the schooling taken at school. Thus, the solution's analytic continuation is possible back in the times when the persons were studying at schools. This is one more important result in my study.

Here, I put down Fisk PDF's statistic moments expressed in terms of the model's parameters; these can be of use in studying age-wage profiles.

$$\text{Mode} = \left(Q \frac{P-1}{P+1}\right)^{\frac{1}{P}} = t_{\max} \tag{17}$$

$$\text{Median} = Q^{\frac{1}{P}} = \left(\frac{P+1}{P-1}\right)^{\frac{1}{P}} t_{\max} \tag{18}$$

$$\text{Mean} = \left(\frac{P+1}{P-1}\right)^{\frac{1}{P}} \frac{\pi}{P} \csc\left(\frac{\pi}{P}\right) \cdot t_{\max} \tag{19}$$

$$Variance = \left(\frac{P+1}{P-1}\right)^{\frac{2}{P}} \left\{ \frac{2\pi}{P} \csc\left(\frac{2\pi}{P}\right) - \left[\frac{\pi}{P} \csc\left(\frac{\pi}{P}\right)\right]^2 \right\} \cdot t_{max}^2 \qquad (20)$$

(Note, moment $n$ exists if $P > n$.)

Rewriting (14) by using (18) makes it handier for studying.

$$C(t) = \frac{2P}{A} \cdot \left\{ \frac{P}{t_{max}} \cdot \frac{\left(\frac{t}{t_{max}}\right)^{P-1}}{\left[1 + \frac{P+1}{P-1}\left(\frac{t}{t_{max}}\right)^P\right]^2} \right\}. \qquad (21)$$

Here, it is interesting to point if $t >> Q^{1/P} > t_{max}$ then (14) has approximation as

$$C(t)\big|_{t >> t_{max}} \sim \frac{1}{t^{P+1}}. \qquad (22)$$

(Compare this to (weak) Pareto law!). I present this formal result here just to stress that (14) and, therefore, (8) make an economic sense.

It was found by many that the age-wage profiles are usually bell-shaped and positively skewed. Weak Pareto law approximates quite well those fattened right tails seen in age-wage profiles (Krämer & Ziebach, 2003). Therefore, to have better results many of researchers apply two laws in data fitting. In the very beginning and in the middle of ranges of age-wage profiles they apply some bell-shaped PDF, say BETA-family, and then Pareto PDF. The problem is persisting in how to mate smoothly these two. The law (14) draws out both bell-shaped region and fattened right tail of the curve. It is the third important result of the study.

I have to note that if there in Eq. (5) $0 > A > -(1 + Z'(1))^2 e^{-Z(1)}$ then $Q < 0$, and the formulas (7) and (8) will be as following.

$$C(t) = \frac{2P^2}{A \cdot Q} \frac{t^{P-1}}{\left[1 - \frac{t^P}{|Q|}\right]^2}, \qquad (23)$$

$$W(t) = \gamma \cdot \left( \frac{2P^2}{A \cdot Q} \frac{t^{P-1}}{\left[1 - \frac{t^P}{|Q|}\right]^2} \right). \qquad (24)$$

Thus, in this case the solution of Eq. (3) does not present itself Fisk PDF. The curve (of type II), (24) being concave at the beginning, then goes up fast, and then, after passing through the break point at $t = |Q|^{1/P}$, goes down (see Fig. 3 in top of the next page).

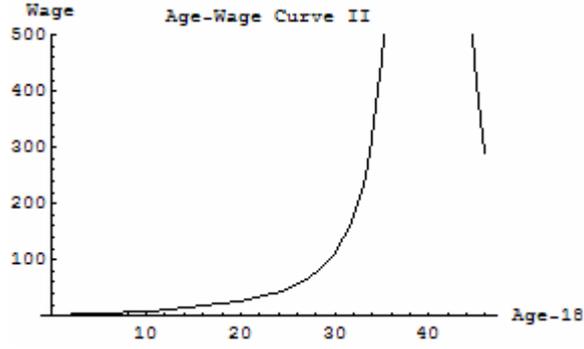

**Fig. 3.** The age-wage profile in the case of (24) (overinvestment) and *γ=1.0*
(*α=0.0036, β=1.5, Z(1)=0.0, Z'(1)=0.87: A=-0.007, P=1.87, Q=-969.4*)

Such effect (some extra peak at the end of the life cycle of wages) has been seen in the studies (see also in Fig. A1 – A13 in the Appendix). Opinions differ widely on why it is so. Having in mind formula (24), it might be explained by an overinvestment in schooling; one can expect there is a (small) cohort of workers of senior ages who do invest in schooling by this strategy. Such pattern of behavior of the cohort should be argued. My explanation is that some intelligent workers risk doing so as they wish to win economic contest with others. Moreover, if the wages are high in a senior age and the worker's wellbeing is high, extra schooling or self-education is among few attractive sectors for investing money. At the end of life cycle of wages, workers in mass being parents decide as a rule to invest money in further schooling of children; those who make alternative decision (e.g., the family have no children) might go on investing in themselves. Those extremely high levels of wages (incomes) predicted by Eq. (24) (at a neighborhood of $t = |Q|^{1/P}$) but not seen in the age-wage profiles are most likely to be suppressed by market mechanisms. It might be the same in the case of varying the coefficients in Eq. (1) in the life cycle of wages. The problem of varying coefficients in Eq. (1) will be discussed in forthcoming article.

**The analysis and experimental proof for the model**

In this section, I apply regression technique to check the model upon reliability. To do this I use formula (see (8))

$$W(t) = \gamma \cdot \left( \frac{2P^2}{A \cdot Q} \frac{t^{P-1}}{\left[1 + \frac{t^P}{Q}\right]^2} \right), \quad (25)$$

and assume that the "true" age-wage profile paid by labor market is

$$\hat{W}(t) = \delta + W(t), \quad (26)$$

where $\delta$ (starting wage) and $\gamma$ in Eqs. (25), (26) are unknown parameters (of linear regression). (Actually, starting wages can be eliminated in data samples by simple subtracting).

Fitting (26) to a data sample is a three-stage process. First, I apply the trial-and-error method to get the best correlation of a data sample on the formula (25) by setting its parameters by hands. Second, I apply linear regression of the data sample on the formula (25) specified by the found parameters to estimate the model's external parameters $\delta$ and $\gamma$. Finally, the parameters are varying a little to get the best regression fit.

To start doing so in line with my approach I need the data on the workers' wages but for the workers of the very same generation and for about 40 years. I could not find such data, and it is very likely there are no such data at all. (If there were such longitudinal study it would be very expensive.) Moreover, if such data do exist, somebody could say these data were of little use; one would argue that neither economy, nor education would be stable for about 40 years.

The available data in age-wage profiles for workers are cross-section ones, i.e. these have been measured on overlapping generations. In the framework of my approach, such data cannot be considered "true". Nevertheless, I use such particular data in regressions. Should be explained, why it is possible.

The new Keynesian economics give us the reasons why the adjustment of wages and prices throughout the economy is staggered. The staggered setting of individual wages makes the overall level of wages sticky (see e.g., Mankiw, 1992). The new Keynesians argue that the stickiness and slowness of change of the wages result also from acting in the markets such factors as contractual work, influence of trade unions, etc. Thus, one can expect that the starting slopes of the age-wage curves may vary but not much in the long run. The workers belong to different generations but they all are working at present time, and therefore their economic behavior is of a general pattern. In the present context, this means that those not observable age-wage profiles at micro (personal) level are parallel to those that are seen in overlapping generations. Therefore, one can speculate that a kind of *ergodicity hypothesis* is true in this case, and thus the macro model (1) is true. The hypothesis is in line also with the approach; averaging, which is used in the model (see e.g. (3)), also suppresses "vibrations" or "noise" in the competencies.

The expression (26) has the form, which is alike Keynes law for aggregate consumption. There, in (26) the first parameter might be considered as workers' "autonomous" wage "assigned" by labor market; the second parameter might be considered as employers' marginal propensity to pay workers for their competencies. Market's mechanisms make the overall level of "autonomous" wages adjust slowly in the long run; employers in mass are also inflexible in changing in the long run of the rates of paying for labor. As for $\alpha$ in Eq. (1), it is quite clear that that is related closely to the psycho-physiological abilities of humans; the humans' abilities in mass (e.g. to acquire and to keep holding the knowledge) do not vary markedly for centuries.

For the testing of the model (1), I used the data on the age-wage profiles for Finnish workers (Elo & Salonen, 2004). The *ordinary least square method* (OLS) was used in the study for fitting curve (25) to the data samples. In the Appendix, I present all the results of modeling. These results and some principal outcomes of the approach are discussed in the next section.

The presented analysis would be considered incomplete if nothing were said about the relevance of Mincer's formula to the solution found.

By expanding (26) to power series, one can write

$$\ln(\hat{W}(t)) = \ln(\delta) + \ln\left[1 + \frac{1}{\delta}\frac{\frac{\gamma}{\delta}C(t)}{(1+\frac{\gamma}{\delta}C_{max})}\right] \cong a_0 + a_1(t-t_0) + a_2(t-t_0)^2 + \dots \quad (27)$$

The unknown parameters $a_i$ then can be found by differentiation (27) with respect to time, and then estimated in every point of the interval *[0, T]*. I found the expressions for $a_i$ in the terms of model's parameters but they turned out to be very sophisticated, and I was not able to give them an economic sense. It was also found, on the grounds of estimates, that the series in the right-hand side of (27) converges too slowly to be effective for the approximation of the left-hand side of (27), and at any point of *[0, T]*. Therefore, I applied OLS to expression

$$\min_{\{a_0, a_1, \ldots, a_m\}} \int_1^T \left[ \ln(\hat{W}(t)) - \sum_{i=0}^{m} \hat{a}_i t^i \right]^2 dt \qquad (28)$$

to calculate the estimates of $\hat{a}_i$ parameters. I could not solve this problem analytically and applied direct polynomial regression to estimate $a_i$ parameters in (28). In the case of $P = 1.872$, $Q = 1925.3$ (see row "Men 2001" in Table A1 in the Appendix), and $m = 2$ I found that $\hat{a}_0 = 9.43989$, $\hat{a}_1 = 0.0622249$, and $\hat{a}_2 = -0.00105059$. The calculated parameters $\hat{a}_i$ have "right" signs according to Mincer's formula. The graphs in Fig.4 show that the quadratic approximation understates the earnings growth at the beginning of the life cycle and overstate the decline in the earnings towards the end of the life cycle. The quadric ($m = 4$) approximation with $\hat{a}_0 = 9.21378$, $\hat{a}_1 = 0.129816$, $\hat{a}_2 = -0.00551125$, $\hat{a}_3 = 0.000108698$, and $\hat{a}_4 = -8.50389E(-7)$ fits to $\ln(\hat{W}(t))$ better; it also fits better (with $\hat{a}_0 = 9.10936$, $\hat{a}_1 = 0.167474$, $\hat{a}_2 = -0.00899044$, $\hat{a}_3 = 0.000224172$ and $\hat{a}_4 = -2.10003E(-6)$) the logarithm of the data. Thus, my model fits the data better than Mincer's, but worse than Murphy & Welch's does.

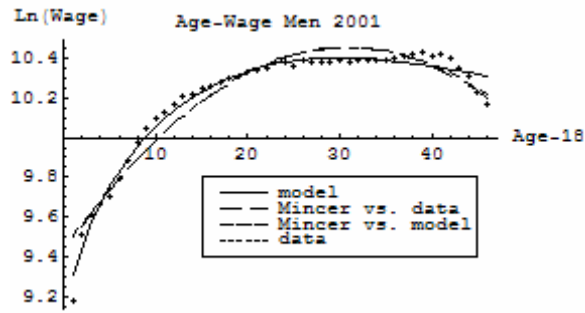

**Fig. 4.** The profile and the best fitting curves of age-wage for Finnish male workers in 2001.

Thus, I do speculate after my study that Mincer earnings function is just an approximation for the exact solution (26) of the problem. The other side of the coin is that the exact solution is specified by the parameters, which are few and they are meaningful. While the formulas (25) and (26) present the regular '"law" made of Eq. (1), Mincer "law" is just an empirical one. As for extension of Mincer formula by simply adding the extra members, any smooth data can be fitted better and better by applying power series of higher order.

## Summary and some open questions

The presented approach is based on the idea that worker's competency has an economic value. The approach also uses the ideas on the motives and patterns of economic behavior clearly expressed by Gary S. Becker in his famous lecture (Becker, 1993). The approach resulting in the novel model takes into account psycho-physiological factors of humans concerning extra schooling and on-the-job training. Thus, both the approach and the model are more relevant than the known ones to consider education factors and humans' abilities to the schooling influencing earnings.

The model resulting in the analytic solution provides clear and simple explanation of mechanisms forming age-wage profiles for workers. These mechanisms provide for endogenous growth of initial competency of worker, worker's competency damping, and changing rate of investment in extra schooling depending on workers' earnings.

There are three important results of my study. I found that the overall process, which forms age-wage profiles, obeys a special case of the Fisk PDF. I found that the solution's analytic continuation is possible back in the times when people who are now the workers were studying at schools. I found that the solution gives quite good fitting to the samples of the observations. Thus, the study can be a ground for further development of the theory of human capital.

Let us discuss in brief some principal outcomes of the study. As for endogenous growth of worker's initial competency, Table A1 in the Appendix shows that condition, $Z'(1) \gg 2A$ pre-assumed for analytic continuation of the solution back in the times of schooling holds. Formally, putting $\alpha = 0$ and $\beta = 0$ in Eq. (1), one can see the endogenous growth of worker's initial competency is set of being linear with time so Mincer formula holds with the model. Thus, both the starting slope and the initial point of solution's curve can be considered a result of schooling the workers had had to that moment when they entered labor market.

The obtained regression curves (see in the Appendix) are very close to those seen in the experimental age-wage profiles for Finnish workers. In all cases, these curves provide for better fitting to the data than the minceranian curves do. The best-fit parameters for the model are listed in Table A1. One can see there, the damping parameters are very stable but there are some gender differences in the damping of competency. These differences are very small in magnitude. I am skeptical about male and female differences in the ability of upholding a competency level. The effect resulting in the differences is rather in the patterns of social behavior; women pay more attention to the families so female workers are some less motivated to keep doing well with professional knowledge than men do. On the other hand, female workers in Finland used to invest in extra schooling more than men did, and the model "says" that women as a rule had more progress in schooling at high schools so their initial competencies were higher than men had. The male workers' initial competencies were paid less than the female workers' ones were. On the other hand, the employers' marginal propensity to pay for the male worker's competency was higher. Thus, men were experiencing discrimination in starting wages, women in the rate of change of wages. These discrimination effects not resulting in the model's nature change were externalities of labor markets, namely, in Finland in the years of 1996, 1985, and 2001.

Thus, the model can provide an easy and meaningful analysis of outcomes of economic behavior of agents in labor market, namely, the age-wage profiles for workers, including motives, patterns, and outcomes of their interacting with an educating environment.

The meanings of the model's inner parameters are as following. The *α*-parameter presents an average rate of damping of competency (depends on the psycho-physiological characteristics of human brain), *β*-parameter is an average rate of investment in extra schooling (depends on the agents' rational expectations), *Z(1)* is the logarithm of initial competency (indicates a level of), and *Z'(1)* is an average rate of (endogenous) growth of the initial competency. One can assume the first two depend on time. Then, the *α*-parameter is likely to follow the "law" for aptitude curves (see e.g. (Phelps J. A., Brookhouse K., 1994)). As for *β*-parameter, I guess that it may depend on time more tangibly as its nature is closely related to the process of decision-making. Rather its behavior is an adaptive one.

The *δ*-parameter and *γ*-parameter are outer ones to the model; these are set by the market. It is interesting to point here, if the inner parameters, the starting wages and the maximum wages are known in advance, then *γ*-parameter can be easily evaluated (see formulas (6a), (6b), (12), and (26)) by the formula:

$$\gamma = \frac{(\hat{W}_{max} - \hat{W}_{start}) \cdot t_{max} \cdot 2A}{Z'(1)^2 + 2Z'(1) + 2A}. \qquad (29)$$

I have to stress once again that the presented model is neither a "clone" of Mincer's nor a type of model that uses ideas of production of human capital and/or allocation of (non-leisure) time (Ben-Porath, 1967). Worker's competency $C(t)$ underlying the approach is used here as a formal function for better visualization of wages $W(t)$. Thus, from an economic standpoint the age-competency profiles for workers are by-product of the approach, but they are not by-product with respect to education. I consider the age-competency model as an effective tool to the study of educational outcomes. To be more accurate with wording "worker's competency', I have to point out here that it has the meaning of "paid competency". In the framework of the approach, it could be named "$E^3$-competency", i.e. "education × economy × earnings & competency".

In conclusion, I have to point out that the presented approach gives theoretical grounds for posing and solving the problems (e.g., optimization) in the field of economics of education, income and lifetime analysis. The approach resulting in the model is based on few ideas and these are quite clear and logical. The analytic solution brought forward with the model can be practically applied to the studying of age-wage profiles for workers. Improvements can be made with the model by assuming its parameters depending on time. Fitting the model's solution to more sample data would result in estimates that might be used for strengthening my theory and for forecasting. (Given the obtained estimates (see Table A1 in the Appendix), I present a forecast for the age-wage profiles for Finnish workers (men) in the year of 2015.) These and other relevant problems will be discussed in forthcoming articles.

# Appendix

**Table A1.** The model's best fitting parameters.

| Cohort (Finnish workers) | The model's parameters | | | | | | | | | | Overall fitting quality | |
|---|---|---|---|---|---|---|---|---|---|---|---|---|
| | Internal | | | | External | | Component | | | | | |
| | $\alpha$ | $\beta$ | $Z(1)$ | $Z'(1)$ | $\delta$ | $\gamma$ | $k_C$ | $A$ | $P$ | $Q$ | Corr. | $R^2$ |
| Men 2001 | 0.0036 | 0.010 | 0 | 0.87 | 8879.91 (525.41) | 2096.28 (54.95) | 0.86 | 0.00364 | 1.872 | 1925.3 | 0.985 | 0.970 |
| Women 2001 | 0.0038 | 0.021 | 0 | 1.04 | 6867.78 (349.90) | 991.64 (26.83) | 1.02 | 0.00388 | 2.042 | 2146.32 | 0.984 | 0.968 |
| Men 1985 | 0.0036 | 0.016 | 0 | 0.93 | 12573.83 (393.40) | 1639.36 (36.28) | 0.92 | 0.00366 | 1.932 | 2038.28 | 0.989 | 0.978 |
| Women 1985 | 0.0038 | 0.021 | 0 | 1.09 | 12022.80 (145.77) | 483.46 (10.23) | 1.07 | 0.00388 | 2.092 | 2252.72 | 0.990 | 0.981 |
| Men 1966 | 0.0036 | 0.043 | 0 | 1.14 | 17986.80 (430.0) | 974.40 (27.20) | 1.09 | 0.00376 | 2.148 | 2436.82 | 0.983 | 0.967 |
| Women 1966 | 0.0038 | 0.620 | 0 | 0.30 | 3638.57 (500.57) | 7773.56 (285.0) | 0.12 | 0.01000 | 1.307 | 339.99 | 0.972 | 0.944 |

Notes: Standard deviations are in parentheses. Corr. = coefficient of linear correlation. $R^2$ = coefficient of determination.

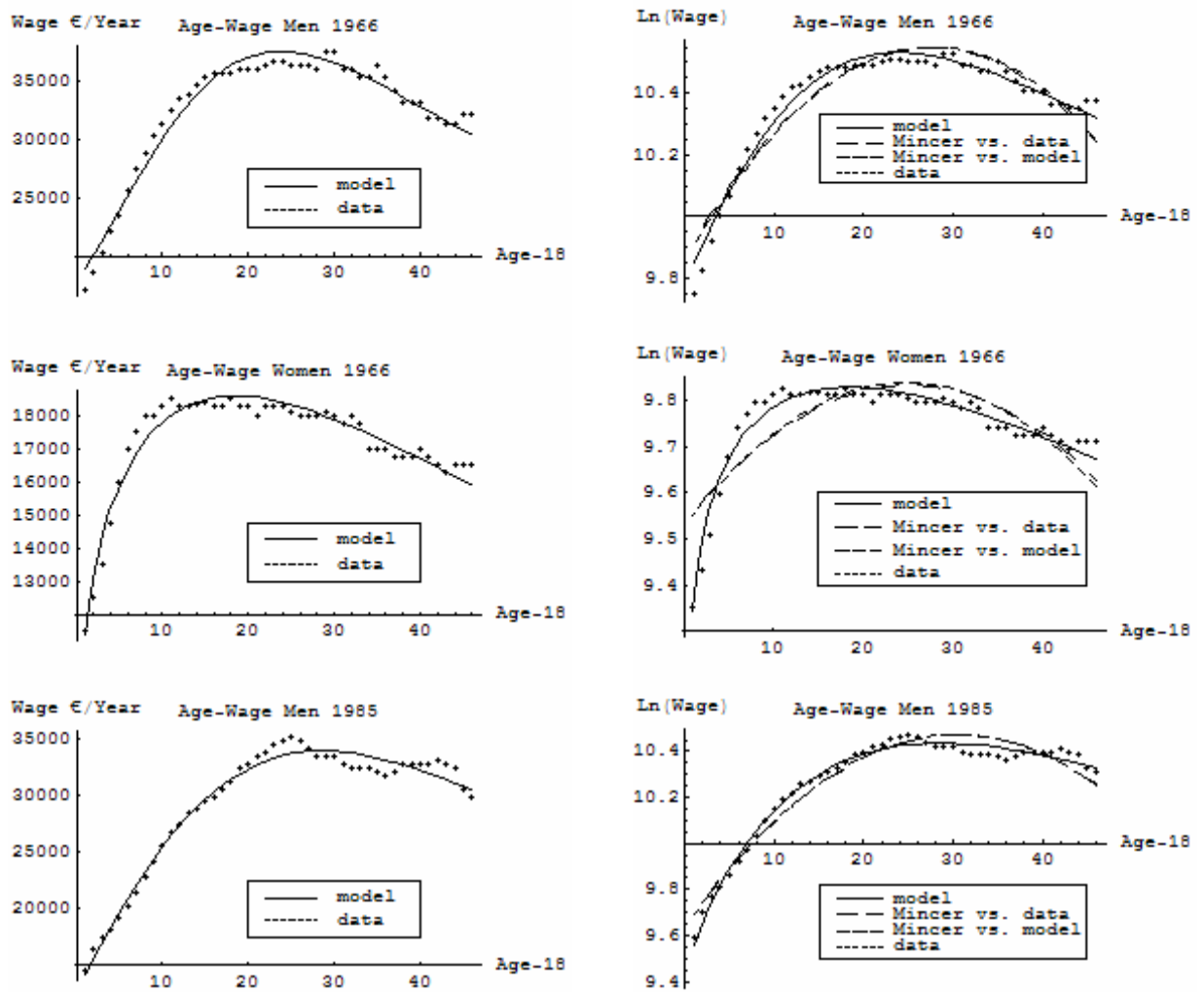

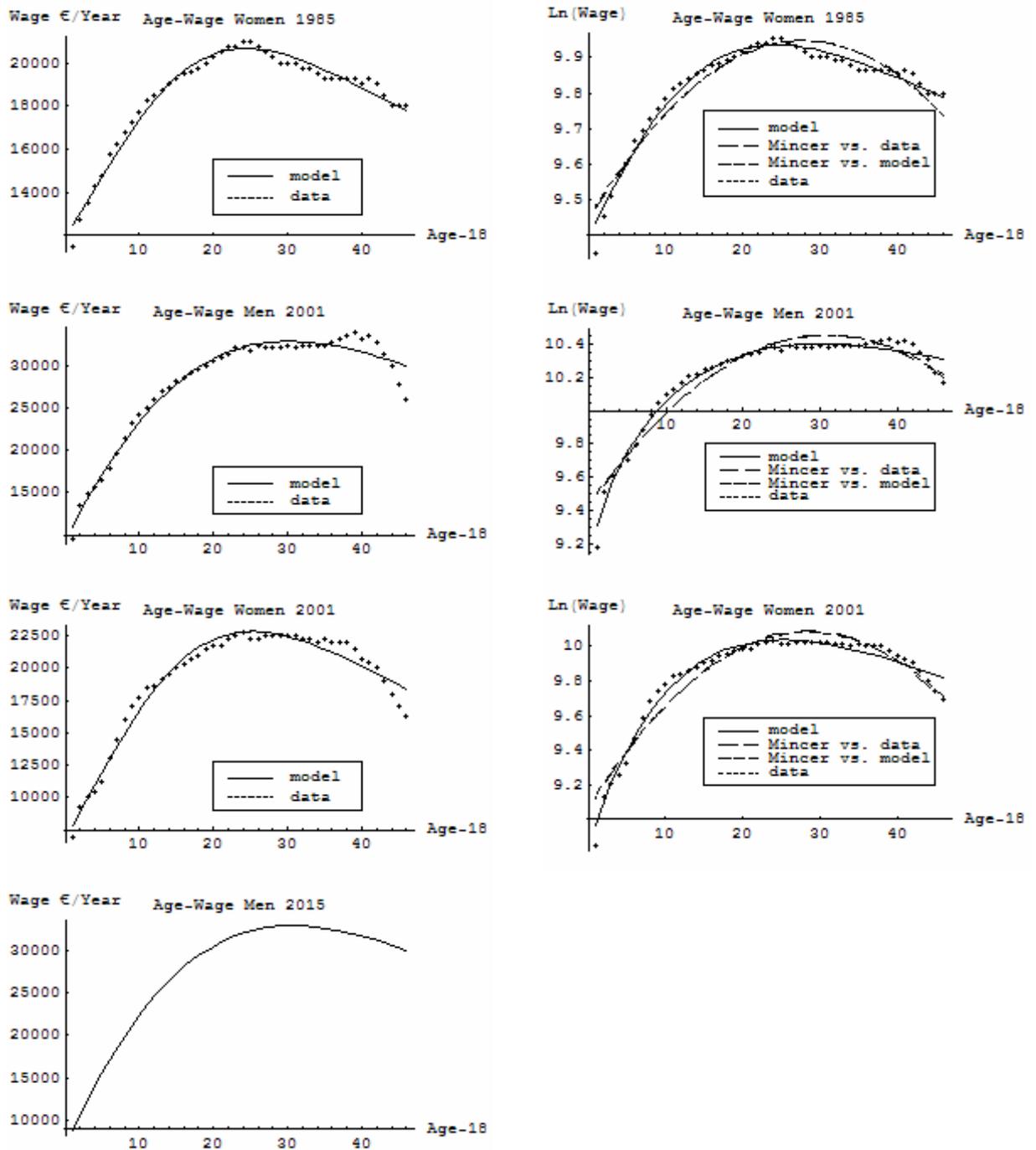

**Fig. A1 – A13.** The profiles and the best fitting curves of age-wage for Finnish workers.

Notes: The source of the data is Elo K., Salonen J. "Age-Wage Profiles for Finnish Workers." Working Papers 7. Finnish Centre for Pensions. Helsinki, 2004. The observed changes in the parameters in Table A1 are underlying the last plot (a forecast for the year of 2015).